\documentstyle[sprocl]{article}

\input{psfig}
\def\be{\begin{equation}}
\def\ee{\end{equation}}
\def\bea{\begin{eqnarray}}
\def\eea{\end{eqnarray}}
\begin{document}

\title{FERMI AND NON--FERMI LIQUID BEHAVIOR OF LOCAL MOMENT SYSTEMS
WITHIN A CONSERVING SLAVE BOSON THEORY}

\author{JOHANN KROHA and PETER W\"OLFLE}

\address{Institut f\"ur Theorie der Kondensierten Materie,
Universit\"at Karlsruhe, Postfach 6980, 76128 Karlsruhe, Germany}


\maketitle\abstracts{The question of Fermi liquid vs. non--Fermi
liquid behavior induced by strong correlations is one of the
prominent problems in metallic local moment systems. As standard
models for such systems, the SU(N)$\times$SU(M) Anderson impurity
models exhibit both Fermi liquid and non--Fermi liquid behavior,
depending on their symmetry. Using an auxiliary boson method,
we present a generally applicable scheme to select the relevant
contributions in the low frequency regime, while preserving the
local gauge symmetry of the model. It amounts to a conserving
T--matrix approximation (CTMA) including coherent spin flip as
well as charge fluctuation processes, which are found to 
dominate in the Kondo and in the mixed valence regime, respectively.
The infrared threshold exponents of the auxiliary particle spectral
functions are indicators for the presence of Fermi or non--Fermi
liquid behavior in any given model with strong on--site repulsion.
We show that, in contrast to earlier auxiliary boson theories,
the CTMA recovers the correct exponents in both cases, indicating
that it correctly describes both the Fermi and the non--Fermi
regimes of the Anderson model.
}

\section{Introduction}
It is a remarkable feature of interacting, itinerant 
fermion systems that at low temperatures $T$ they behave in 
general in much the same way as a noninteracting Fermi gas,
even though the interaction may be strong. An extremely successful
description of this phenomenon,
known as Fermi liquid (FL) behavior, is provided  
by the notion of quasiparticles, which was established
by Landau's phenomenological Fermi liquid theory \cite{landau.53}. 
The key assumption is that, as the interaction is continuously
turned on, there exists a 1:1 correspondence 
between the low energy eigenstates of the interacting system and
the single--particle states of the free Fermi gas. Therefore,
the low--lying interacting states may be described approximately
as single--particle states or quasiparticles, whose decay rate
$1/\tau$ is small compared to their excitation energy $\omega$,
$1/\tau \ll \omega$, and which are characterized by the 
same quantum numbers as the noninteracting states. 
As a consequence, Fermi liquids exhibit the
same low--$T$ thermodynamics as a noninteracting Fermi system,
e.g.~a linear in $T$ specific heat $c=\gamma T$ and a
constant Pauli paramagnetic susceptibility $\chi _o$. However,
the effective mass and other parameters may be renormalized by
the interaction, resulting in an enhancement of the 
specific heat coefficient $\gamma$ and the susceptibility 
$\chi _o$. It is at the heart of the quasiparticle picture
that at low $T$ the Pauli exclusion principle substantially
reduces the phase space available to quasiparticle scattering.
This blocking mechanism is effective as long as the 
quasiparticle interaction is {\it shortranged in space and time}, 
which is usually the case in three dimensions because of screening.
It also implies that the quasiparticle scattering rate
vanishes as $1/\tau \propto (\omega ^2 + T^2)$, thus providing
a microscopic justification for the basic assumption of FL theory
and leading to an interaction contribution to the electrical
resistivity which behaves as $\Delta \rho \propto T^2$. 
Obviously, the Pauli principle as the origin of FL behavior is
very robust, which explains the almost ubiquitous presence of
a FL ground state in interacting Fermi systems and the broad
success of Fermi liquid theory.

In this light it is all the more exciting that in recent years
a number of new alloys have been discovered which exhibit 
striking deviations from this usual behavior. Among these, an
important subclass are certain heavy fermion compounds on the
basis of Ce$^{3+}$ or U$^{4+}$ ions, e.g.~CeCu$_{6-x}$Au$_x$ 
\cite{loehn.94,loehn.96}, CeCu$_2$Si$_2$ \cite{steglich.96}, 
CePd$_2$Si$_2$ \cite{lonzar.96a,lonzar.96b} and Y$_{1-x}$U$_x$Pd$_3$
\cite{maple.94} or 
UCu$_{5-x}$Pt$_x$ \cite{maple.96}. In these materials logarithmic
or fractional power law deviations from FL behavior have been
observed in their thermodynamic as well as transport properties
at low temperatures. These systems have in common that a localized,
degenerate degree of freedom, the magnetic moments of the Ce or
U ions, is dynamically coupled to a continuum of conduction
electron states. In general, such a coupling generates the Kondo
effect, characterized by resonant spin flip scattering of electrons 
at the Fermi surface off the local moment. Concomitantly, the
conduction electron spin flip rate initially increases 
logarithmically as the temperature is lowered, passes 
through a maximum at a 
characteristic scale, the Kondo temperature $T_K$, and approaches 
zero as $T\rightarrow 0$, because the effective local moment 
becomes screened by the conduction electron spins. 
Thus, even for many strongly correlated systems of this type a 
Fermi liquid description applies below $T_K$, with usually a strongly 
enhanced quasiparticle effective mass, lending the term
``heavy fermion systems'' to these materials. 

Completely new physics may arise, however, if the quenching of 
the local moments is inhibited. Two different mechanisms for this 
to occur have been put forward:
\begin{itemize}
\item[(1)] Proximity of a quantum phase transition (QPT) to an
antiferromagnetically ordered state 
\cite{millis.93}$^-$\cite{sachdev.95}
as a function of a dopant concentration $x$ or of pressure.
Near the QPT the quantum critical fluctuations become 
longranged in space and time and can, thus, mediate a longrange
quasiparticle interaction, leading to a breakdown of FL
theory. There are indications for this spatially extended 
mechanism to be realized near the QPT of the Ce based 
compounds 
\cite{loehn.94}$^-$\cite{lonzar.96b}.
\item[(2)] Two--channel Kondo effect (2CK) \cite{nozieres.80,coxzawa.98}. 
The local
magnetic moment is coupled to two exactly degenerate conduction
electron channels. Because of a frustration effect between 
the screening of the local moment by the different conduction
channels the moment quenching cannot be complete, leading to 
a nonvanishing conduction electron spin scattering rate even 
at the lowest temperatures and
subsequently to a breakdown of FL behavior. It has been 
suggested \cite{cox.87} that this mechanism, based on single--ion
dynamics rather than longrange fluctuations, may be
realized predominantly in the U based materials with cubic 
symmetry about the magnetic ion, which do not exhibit a QPT.
\end{itemize}
\noindent

While both mechanisms provide possible pathways to non--FL
behavior, neither one can at present consistently explain the 
wealth of experimental data showing non--FL behavior at 
low temperatures. Open questions in the
QPT picture include, e.g., whether the local impurity
dynamics competing with the magnetic ordering can play a role,
and how the transition from the spin screened heavy FL phase to the
magnetically ordered phase occurs alltogether.
In the 2CK mechanism, on the other hand, inter--impurity 
interactions could modify the single ion behavior. 
%Thus,
%a complete understanding of the non--FL systems at hand and the 
%nature of their elementary excitations is far from being reached.
Exact solution methods as well as numerical simulations have 
provided important progress in our understanding of
strongly correlated quantum impurity systems. 
However, their applicability is essentially restricted 
to problems involving only a single 
impurity, owing to integrability conditions or limitations
in the numerical effort, respectively. Therefore, more generally 
applicable theoretical techniques are called for. In the present
work we focus on the single--ion dynamics. 
We develop a standard field theoretical method,
based on an auxiliary particle or slave boson representation, 
which describes 
the quantum impurity dynamics in a controlled way and 
at the same
time has the potential of being extended to problems of many
impurities on a lattice. As a standard model of strongly
correlated electrons which, depending on its symmetry, 
exhibits both FL and non--FL behavior, we consider the 
SU(N)$\times$SU(M) 
Anderson impurity model of a local, $N$--fold degenerate degree of
freedom, coupled to $M$ identical conduction bands.

In order to set the stage for the more formal development of the
theory, in the following section we will briefly review the
striking differences in the phenomenology of the single--channel
and the multi--channel Kondo effects. In section 3 the slave boson 
representation is introduced, which provides a particularly 
compact formulation of the SU(N)$\times$SU(M) Anderson model.
We also discuss why the presence of FL or non--FL behavior in a
given quantum impurity system can already be seen from the singular
infrared dynamics of the auxiliary particles.
Section 4 contains a critical assessment of earlier approximate
slave boson treatments. 
This will motivate our approach of conserving
slave boson approximations, which is developed in section 5.
As will be seen, the results produced from this  
theory are in very good agreement with known exact properties of 
the model. Conclusions are drawn in section 6.

\section{Single-- and multi--channel Kondo effect and physical realizations}
\begin{figure}
\vspace*{-0cm}
\hspace*{0.5cm}{\psfig{figure=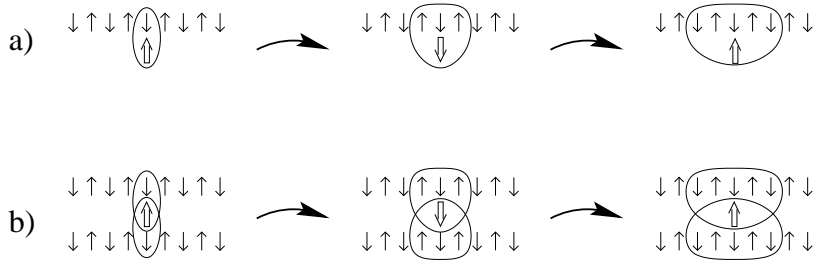,width=10.7cm}}
\caption{
Sketch of the renormalization group for a) the single--channel Kondo model
(local moment compensation) and 
b) the two--channel Kondo model (local moment over--compensation). 
Small arrows  denote conduction electron spins 1/2, a heavy
arrow a localized spin 1/2. The curved arrows indicate successive 
renormalization steps.
\label{cartoon}}
\end{figure}

In this section we briefly discuss how the single--channel and the 
two--channel Kondo effects may arise in magnetic metals, if the interaction
between the local moments is weak.  
A local moment is generated by an atomic f or d level
whose energy $E_d$ lies far below the Fermi energy $\varepsilon _F \equiv 0$ 
and whose electron occupation number is effectively restricted to 
$n_d  \le 1$ by a strong Coulomb
repulsion $U$ between two electrons in the same orbital. While the
angular momentum degeneracy of the level is usually lifted by crystal field
splitting, a twofold degeneracy of a level occupied by one electron 
is guaranteed by time reversal symmetry (Kramers doublet) in the absence
of a magnetic field, corresponding to 
the spin quantum numbers $m=\pm 1/2 $ of the electron. In addition, there
is a hybridization matrix element $V$ between the atomic orbital and the
conduction electron states. Such a system is described by the
single impurity Anderson hamiltonian
\begin{equation}
H=\sum _{\vec k,\sigma }\varepsilon _{\vec k}
c_{\vec k\sigma}^{\dag}c_{\vec k\sigma}+
E_d\sum _{\sigma} d_{\sigma}^{\dag}d_{\sigma}+
V\sum _{\vec k,\sigma }(c_{\vec k\sigma}^{\dag}d_{\sigma}
+h.c.) + 
U d^{\dag}_{\uparrow}d^{\phantom{\dag}}_{\uparrow}
d^{\dag}_{\downarrow}d^{\phantom{\dag}}_{\downarrow} ,
\label{Ahamilton}
\end{equation}
where $c^{\dag}_{\vec k \sigma}$ and $d^{\dag}_{\sigma}$ are the creation
operators of a conduction electron with dispersion 
$\varepsilon _{\vec k}$ and of an electron in the local orbital with spin
$\sigma $, respectively. 
The low energy physics of this system is dominated by processes of second 
order in $V$, by which an electron hybridizes into the
conduction band and the impurity level is subsequently filled by another
electron, thereby effectively flipping the impurity spin.
Thus, in the region of low excitation energies, the Anderson hamiltonian
(\ref{Ahamilton}) may be mapped  
onto the s--d exchange (or Kondo) model \cite{schriwo.66}, 
the effective coupling between the impurity spin and the 
conduction electron spin always being
antiferromagnetic: $J=|V|^2/|E_d| > 0$ ($U\gg |E_d|$). 
These models have been studied extensively by means
of Wilson's renormalization group \cite{wilson.75} and by the Bethe ansatz
method \cite{andrei.83}. In this way the following physical picture has 
emerged.
Resonant spin flip scattering of electrons at the Fermi surface 
off the local degenerate level leads to logarithmic contributions to the 
magnetic susceptibility, the linear specific heat coefficient and 
the resistivity, $\chi(T),\ \gamma (T),\ \Delta\rho (T) 
\propto {\rm ln}(T/T_K)$,
for $T \buildrel >\over\sim T_K$, 
and to a breakdown of perturbation theory at the Kondo temperature 
$T_K = D (N {\cal N}(0)J)^{(M/N)}{\rm exp}\{ -1/(N {\cal N}(0)J)\}$. Here
${\cal N}(0)$ and $D$ denote the density of states at the Fermi energy and the
high energy band cutoff, respectively, and $N$, $M$ are the degeneracy
of the local level and the number of conduction electron channels 
(see below). Below $T_K$ a collective many--body spin singlet 
state develops in which the
impurity spin is screened by the conduction electron spins as lower and
lower energy scales are successively approached, leaving the system with
a pure potential scattering center. The spin singlet formation  
is sketched in Fig.~\ref{cartoon} a) and corresponds to a 
vanishing entropy at $T=0$, $S(0)=0$. It also leads to saturated behavior of
physical quantities below $T_K$, like $\chi (T)=const$, $c (T)/T =const.$ 
and $\Delta \rho (T)\propto T^2$, i.e. to
Fermi liquid behavior.

As an example of possible two--channel Kondo systems we discuss  
the uranium based compounds mentioned in the introduction.
The U$^{4+}$ ions have nominally a
5f$^2$ configuration, i.e. an even number of electrons, which does not
allow for a Kramers degenerate ground state because of integer total spin.
However, in the cubic crystal symmetry of these materials 
the orbital degeneracy may be not completely lifted, so that there can be an 
approximate twofold degeneracy of the U$^{4+}$ ground state, corresponding to
two different orientations of the electrical quadrupole moment of the 5f
orbital in the lattice (quadrupolar Kondo effect) \cite{coxzawa.98,cox.87}.
This degree of freedom may be flipped by scattering of conduction electrons
(which in the cubic symmetry also have a twofold angular momentum degeneracy).
The conduction electron spin is conserved in this scattering process, leaving
it as a Kramers degenerate scattering channel degree of freedom, which we
will label by $\mu = 1,...,M$, $M=2$. Describing the
orbital degree of freedom as a pseudospin 1/2, labelled by the quantum number
$\sigma = 1,\dots ,N$, $N=2$, in analogy to the magnetic Kondo effect, 
we arrive at the SU(2)$\times$SU(M) symmetric Kondo model, 
\begin{equation}
H=\sum _{\vec k,\sigma,\mu}\varepsilon _{\vec k}
c_{\vec k\sigma}^{\dag}c_{\vec k\mu\sigma}+
J \sum _{\vec k, \vec k ',\sigma,\sigma ',\mu}c^{\dag}_{\vec k \mu \sigma }
\vec S\cdot \vec \tau _{\sigma\sigma '}c^{\phantom{\dag}}_{\vec k \mu \sigma '},
\label{2CKhamilton}
\end{equation}
where $\vec S$ is the local pseudospin operator and 
$\vec \tau _{\sigma\sigma'}$ the vector of Pauli matrices. To keep the naming
uniform, we will refer to the orbital degree of freedom as the (pseudo)spin
or local moment, $\sigma$, in analogy to the magnetic Kondo effect, 
and to the physical electron spin as the channel degree of freedom, $\mu$.
In the multi--channel case, too, the conduction electrons of each channel
{\it separately} screen the impurity moment by multiple spin scattering 
at temperatures below the Kondo scale $T_K$. However, in this case,
the local moment is over--compensated, since 
the impurity spin can never form a singlet state with both conduction 
electron channels at the same time in this way, as can be seen in 
Fig.~\ref{cartoon} b).
As a consequence of this frustration, there is not a unique ground state,
leading to a finite residual entropy \cite{tsvelik.85} 
at $T=0$ of $S(0)=k_B {\rm ln}\sqrt{2}$ in the two--channel model. 
In particular, the precondition of FL theory of a 1:1 correspondence 
between interacting and non--interacting states is violated. 
As a consequence, characteristic singular temperature dependence 
\cite{wiegmann.83,affleck.91}
of physical quantities persists for $T\buildrel <\over\sim T_K$ down to
$T=0$:  $\chi (T)\propto -{\rm ln}(T/T_K)$, $c (T)/T  \propto -{\rm ln}(T/T_K)$ 
and $\rho (T)-\rho(0)\propto -\sqrt {T/T_K}$. Note, however, that this
behavior may be changed by any crystal field splitting of the 
quadrupolar non--Kramers ground state doublet.

In order to 
apply standard field theoretical methods to the multi--channel Kondo model
it is convenient to consider it as the low--energy
limiting case of a corresponding Anderson model as discussed 
for the single--channel case. Here, in addition, 
the conservation of the channel degree of freedom has to be guaranteed. 
This can be implemented in an elegant way using an auxiliary boson
representation to be discussed in the next section.

\section{Auxiliary particle representation} 
As discussed above, the local level of a quantum 
impurity in the limit of infinitely strong local Coulomb repulsion $U$ between
electrons in the same level allows only for at most single electron
occupation of the level, $n_d\le 1$. One should note that for a realistic
finite value of $U$ the low--energy physics of the model is effectively
still confined to the part of the Hilbert space without multiple occupancy.
Therefore, the model Eq.~(\ref{Ahamilton}) in the limit 
$U\rightarrow \infty$ is the generic model for the physics of quantum 
impurities at large $U$ in general. 

A powerful technique for implementing the projection in Hilbert space 
caused by a large Coulomb repulsion $U$ is the method of auxiliary particles
(slave bosons, pseudofermions) \cite{barnes.76}. 
Each Fock state $|\alpha\rangle$ of the
impurity is assigned a creation operator, which can be envisaged
as creating the state out of a vacuum state $|{\rm vac}\rangle$ without
any impurity level at all, 
$|\alpha\rangle = a_{\alpha}^{\dag} |{\rm vac}\rangle$.
(E.g., for a single orbital there are four such states, 
$|0\rangle$ (empty orbital), $|\uparrow\rangle$ or $|\downarrow\rangle$
(orbital occupied by a single electron with spin $\uparrow$ or $\downarrow$)
and $|2\rangle$ (level occupied by two electrons with spin $\uparrow$ and
$\downarrow$).) 
Due to the requirements of Fermi statistics, the creation operators 
$a_{\alpha}^{\dag}$ are 
Fermi (Bose) operators for the states holding an odd (even) number of
electrons (or vice versa). The physical state corresponds to the sector of
Fock space with exactly one auxiliary particle, $\sum _{\alpha}n_{\alpha} =1$,
where $n_{\alpha}=a_{\alpha}^{\dag}a_{\alpha}$ is the occupation number
operator of particles $\alpha$.
Compared to alternative ways of effecting the projection, the auxiliary 
particle method has the advantage of making available the powerful
machinery of quantum field theory, provided the constraint on the total
auxiliary particle number can be incorporated in a satisfactory way.
 
For the quantum impurity models of the Anderson type introduced in the 
preceding section, only particles creating empty and singly occupied states are
needed. We define $N$ pseudofermion creation operators $f_{\sigma}^{\dag}$
for each of the singly occupied states (labelled by $\sigma =1,2,\dots N$)
and $M$ boson creation operators $b_{\mu}^{\dag}$ for each of the empty
states created when an electron hops from the impurity into the $\mu$-th
conduction electron band (labelled by $\mu = 1,2,\dots M$). In terms of these
operators the Hamiltonian of the SU(N)$\times$SU(M) Anderson model 
Eq.~(\ref{Ahamilton}) takes the form
\begin{equation}
H=\sum _{\vec k,\sigma ,\mu}\varepsilon _{\vec k}
c_{\vec k\mu\sigma}^{\dag}c_{\vec k\mu\sigma}+
E_d\sum _{\sigma} f_{\sigma}^{\dag}f_{\sigma}+
V\sum _{\vec k,\sigma ,\mu}(c_{\vec k\mu\sigma}^{\dag}b_{\mu}^{\dag}f_{\sigma}
+h.c.)
\label{sbhamilton}
\end{equation}
In addition, the operator constraint
\begin{equation}
Q\equiv \sum _{\sigma} f_{\sigma}^{\dag}f_{\sigma} +
        \sum _{\mu} b_{\mu}^{\dag}b_{\mu} =1
\end{equation}
has to be satisfied at all times. One might interpret the constraint as
a statement of charge quantization, with the integer $Q$ the conserved,
quantized charge. Similar to quantum field theories with conserved charges,
the charge conservation is intimately related to the existence of a local
gauge symmetry. Indeed, the system defined by the Hamiltonian 
Eq.~(\ref{sbhamilton}) is invariant under simultaneous local $U(1)$ gauge
transformations $f_{\sigma}\rightarrow f_{\sigma} {\rm e}^{i\phi (\tau )}$,
$b_{\mu}\rightarrow b_{\mu} {\rm e}^{i\phi (\tau )}$, with $\phi (\tau )$ an
arbitrary time dependent phase. 

While the gauge symmetry guarantees the 
conservation of the quantized charge $Q$, it does not single out any 
particular $Q$, such as $Q=1$. In order to effect the projection onto the
sector of Fock space with $Q=1$, one may use a procedure first proposed by
Abrikosov \cite{abrikosov.65}: Consider first the grand canonical ensemble
with respect to $Q$, defined by the statistical operator
\begin{equation}
\hat \rho _G = \frac{1}{Z_G} {\rm e}^{-\beta (H+\lambda Q)},
\end{equation}
where $Z_G={\rm tr}[{\rm exp}(-\beta (H+\lambda Q))]$ is the grand canonical
partition function and the trace extends over the complete Fock space,
including summation over $Q$. The expectation value of an observable
$A$ in the grand canonical ensemble is given by
\begin{equation}
\langle A\rangle _G = {\rm tr} [\hat \rho _G A] .
\end{equation}
The physical expectation value of $A$, $\langle A\rangle$, 
is to be evaluated in the canonical ensemble where $Q=1$. It can be 
calculated from the grand canonical ensemble by taking the chemical 
potential $\lambda$ to infinity,
\begin{equation}
\langle A\rangle = \lim _{\lambda \rightarrow\infty}\frac{\langle Q A\rangle _G}
{\langle Q \rangle _G}.
\end{equation}
In the following we will concentrate on the auxiliary particle Green's
functions in the grand canonical ensemble as the basic building blocks
of the theory. In imaginary time representation they are defined as
\begin{eqnarray}
G_{f\sigma}(\tau_1-\tau_2) & = - \langle T \{f_\sigma
(\tau_1)f_\sigma^{\dag}(\tau_2)\}\rangle_G\nonumber \\
G_{b\mu}(\tau_1 - \tau_2) & = - \langle T\{ b_\mu
(\tau_1)b_\mu^{\dag}(\tau_2)\}\rangle_G ,
\end{eqnarray}
where $T$ is the time ordering operator.  The Fourier transforms of
$G_{f,b}$ and of the local conduction electron Green's
function may be expressed in terms of the self-energies
$\Sigma_{f,b,c}$ as
\begin{equation}
G_{f,b,c}(i\omega_n) = \Big\{[G_{f,b,c}^0(i\omega_n)]^{-1} -
\Sigma_{f,b,c}(i\omega_n)\Big\}^{-1}
\label{green}
\end{equation}
where
\begin{eqnarray}
G_{f\sigma}^0(i\omega_n) &=& (i\omega_n - E_d - \lambda)^{-1}\nonumber
\\
G_{b\mu}^0(i\omega_n) &=& (i\omega_n - \lambda)^{-1}\label{green0} \\
G_{c\mu\sigma}^0 (i\omega_n) &=& \sum_{\vec k} (i\omega_n -
\epsilon_{\vec k}) ^{-1}
\nonumber
\end{eqnarray}
The Green's fucntions $G_{f,b,c}$ have the following spectral
representations
\begin{equation}
G_{f,b,c}(i\omega_n) = \int_{-\infty}^\infty \frac{d\omega{'}}{\pi}
\frac{A_{f,b,c}(\omega{'})}{i\omega_n - \omega{'}}.
\end{equation}
The projected Green's functions $G_{f,b}$ are obtained by taking the
limit $\lambda \rightarrow \infty$ as discussed above.  As a
consequence, the energy eigenvalues of $H + \lambda Q$ are shifted by
$\lambda Q$.  It is useful to shift the zero of the frequency scale by
$\lambda$ (in the $Q = 1$ sector) and to define the projected spectral
functions as 
\begin{equation}
{\cal A}_{f,b}(\omega ) = \lim_{\lambda\to\infty} A_{f,b}(\omega +
\lambda)
\end{equation}
At zero temperature the ${\cal A}_{f,b}(\omega )$ have the Lehmann
representation
\begin{eqnarray}
{\cal A}_f (\omega )& = &\sum _n \mid \langle 1,n\mid f_{\sigma}^{\dag}\mid
0,0\rangle \mid^2\delta (\omega + E_0^0 - E_n^1)
\label{lehmann}
%\nonumber \\
%{\cal A}_f^{(-)} (\omega )& = &\sum _n \mid \langle 1,0\mid f_{\sigma}^{\dag}
%\mid 0,n\rangle\mid^2\delta (\omega + E_n^0 - E_0^1)
\end{eqnarray}
and correspondingly for ${\cal A}_b$, where $E_n^0$ are the
energy eigenvalues ($E_0^Q$ is the ground state energy) and $\mid
Q,n\rangle $ the many--body eigenstates of $H$ in the sector $Q$ of Fock
space. 
The ${\cal A}_{f,b}$ show
threshold behavior at $\omega = E_0 \equiv E_0^1 - E_0^0$, with ${\cal
A}_{f,b}(\omega ) \equiv 0$ for $\omega < 0$. 
The vanishing imaginary part at frequencies $\omega < 0$ may be
shown to be a
general property of all quantities involving slave particle
operators, e.g. also of auxiliary particle selfenergies and
vertex functions.
%Both ${\cal A}_{f,b}^{(+)}$ and ${\cal A}_{f,b}^{(-)}$ show
%threshold behavior at $\omega = E_0 \equiv E_0^1 - E_0^0$, with ${\cal
%A}_f^{(+)}(\omega ) \equiv 0$ (${\cal A}_f^{(-)}(\omega ) \equiv 0$) 
%for $\omega < 0$ ($\omega > 0$).

Both ${\cal A}_{f}(\omega)$ and ${\cal A}_{b}(\omega)$ 
are found to diverge for $\omega
\rightarrow E_0$ in a power law fashion (infrared singularity)
\begin{equation}
{\cal A}_{f,b} \sim \mid \omega - E_0\mid
^{-\alpha_{f,b}}\theta (\omega - E_0)
\end{equation}
due to a diverging number of particle--hole excitation processes in the
conduction electron sea for $\omega \rightarrow E_0$.

For the single channel case $(M=1)$, i.e. the usual Kondo or mixed
valence problem, the exponents $\alpha_f$ and $\alpha_b$ can be found
analytically from the following argumentation.  Anticipating that in
this case the impurity spin is completely screened by the conduction
electrons at temperature $T = 0$, leaving a pure-potential scattering
center, the ground state $\mid 1,0 \rangle$ is a slater determinant of one
particle scattering states, characterized by scattering phase shifts
$\eta_\sigma^0$ in the s-wave channel (assuming for simplicity a
momentum independent hybridization matrix element $V$).  The
scattering phase shifts are related to the fraction of conduction
electrons attracted (or repelled) by the impurity, $\Delta n_\sigma$,
via the Friedel sum rule as $\eta_\sigma^0 = \pi \Delta n_\sigma$.
The change in the average number of conduction electrons per scattering 
channel $\sigma $ due to the
presence of the impurity exactly matches the
average occupation number of the impurity level per spin channel, 
$\Delta n_\sigma = -n_d/N$, 
where $n_d$ is the total occupation of the impurity level. 
In the Kondo limit $n_d \rightarrow 1$ and for a spin $1/2$ impurity 
this leads to resonance scattering, $\eta_\sigma^0 = \pi/2$.
To calculate now the spectral function ${\cal A}_f(\omega)$ from
Eq.~(\ref{lehmann}), 
one needs to evaluate $\langle 1,n\mid f_\sigma^{\dag}\mid 0,0
\rangle$, which
is nothing but the overlap of two slater determinants, the ground
state of the conduction electron system in the absence of the
impurity combined with 
the decoupled impurity level occupied by an electron
with spin $\sigma$, $f^{\dag}_{\sigma}|0,0\rangle$, on the one hand, 
and the eigenstates of the fully
interacting Kondo system, $|1,n\rangle$, on the other hand.  
As shown by Anderson \cite{anderson.67}, 
the overlap of the two ground state slater determinants
$\langle 1,0\mid f_\sigma^{\dag}\mid 0,0 \rangle$ tends to zero in
the thermodynamic limit (orthogonality catastrophe).  This implies
that the spectral functions diverge at the threshold as
\begin{equation}
A_{f\sigma}(\omega) = c \mid\omega\mid^{-\alpha_f}
\end{equation}
where $\alpha_f = 1 - \sum_{\sigma{'}}(\eta_{\sigma{'}}/\pi)^2$.
Here the $\eta_\sigma{'}$ are the scattering phase shifts relative to
the state $f_\sigma^{\dag}\mid 0,0\rangle$. Using again the Friedel sum rule one
finds $\eta_{\sigma{'}} = \pi n_d/N$ for $\sigma{'} \not= \sigma$.  In
the scattering channel $\sigma$ the change in the number of conduction
electrons is $\Delta n_\sigma = n_d/N - 1$, since the occupation in
the initial state $n_{d\sigma}^{in} = 1$ is reduced by
interaction processes to $n_{d\sigma} = n_d/N$ in the final state.  
As a result, one finds \cite{mengemuha.88} the exponent
$\alpha_f = (2n_d - n_d^2)/N$. A similar consideration for the slave
boson spectral function yields $\alpha_b = 1 - n_d^2/N$.  These
results have been found independently from Wilson's numerical
renormalization group approach \cite{costi.94} and using the Bethe ansatz
solution and boundary conformal field theory \cite{fujimoto.96}.  
It is interesting
to note that (i) the exponents depend on the level occupancy $n_d$
(in the Kondo limit $n_d \rightarrow 1$, $\alpha_f = 1/N$ and
$\alpha_b = 1 - 1/N$, whereas in the opposite, empty orbital, limit
$n_d \rightarrow 0$, $\alpha_f \rightarrow 0$ and $\alpha_b
\rightarrow 1$) (ii) the sum of the exponents $\alpha_f + \alpha_b = 1
+ 2 \frac{n_d(1-n_d)}{N} \geq  1$.

We stress that the above derivation of the infrared exponents
$\alpha_{f,b}$ holds true only if the impurity complex acts as a pure
potential scattering center at $T = 0$.  This is equivalent to the
statement that the conduction electrons behave locally, i.e. at the
impurity site, like a Fermi liquid. Conversely, in the multi--channel 
(non--FL) case, $N\geq 2$, $M\geq N$, the exponents have been found from
a conformal field theory solution \cite{affleck.91} of the problem
in the Kondo limit to be $\alpha _f = M/(M+N)$, $\alpha _b = N/(M+N)$,
which differ from the FL values. Thus, one may infer from
the values of $\alpha_{f,b}$ as a function of $n_d$, whether or not
the system is in a local Fermi liquid state.

The physical electron Green's functions for the local d-electrons and
for the conduction electrons ${\cal G}_d(i\omega_n)$ and ${\cal
G}_c(\vec k,\vec k{'}\;;\;i\omega_n)$, can be expressed through the
self-energy $\Sigma_c(i\omega_n)$ as \cite{costi.96}
\begin{equation}
{\cal G}_d(i\omega_n) =
\frac{1}{V^2}\lim_{\lambda\to\infty}
e^{\beta\lambda}\Sigma_c(i\omega_n)
\end{equation}
and
\begin{equation}
{\cal G}_c(\vec k,\vec k{'}\;;\;i\omega_n) = G_c^0(\vec
k,i\omega_n)\Big[\delta_{\vec k,\vec k{'}} + V^2 {\cal
G}_d(i\omega_n)G_c^0(\vec k{'},i\omega_n)\Big]
\end{equation}

\section{Mean field and non-crossing approximations}
For physical situations of interest, the $s-d$ hybridization of the
Anderson model (\ref{Ahamilton}) is much smaller than the conduction band width,
$N_0 V \ll 1$, where $N_o = 1/2D$ is the local conduction electron
density of states at the Fermi level.  This suggests a perturbation
expansion in $N_oV$.  A straightforward expansion in terms of bare
Green's functions is not adequate, as it would not allow to capture
the physics of the Kondo screened state, or else the infrared
divergencies of the auxiliary particle spectral functions discussed in
the last section.  In the framework of the slave boson representation,
two types of nonperturbative approaches have been developed.  The
first one is mean field theory for both the slave boson amplitude 
$\langle b\rangle$
and the constraint ($\langle Q\rangle = 1$ rather than $Q = 1$).  The second one
is resummation of the perturbation theory to infinite order.

\subsection{Slave boson mean field theory} 
Slave boson mean field theory is based on the assumption that the slave
bosons condense at low temperatures such that $\langle b_\mu \rangle \neq 0$.
Replacing the operator $b_\mu$ in $H + \lambda Q$  
by $\langle b_\mu \rangle$ (see Ref.~[22]), 
where $\lambda$ is a Lagrange multiplier 
to be adjusted such that
$\langle Q \rangle = 1$, one arrives at a resonance level model for the
pseudofermions.  The position of the resonance, $E_d + \lambda$, is
found to be given by the Kondo temperature $T_K$, and is thus close to
the Fermi energy. The resonance generates the low energy scale $T_K$,
and leads to local Fermi liquid behavior.  While this is qualitatively
correct in the single--channel case, it is in blatant disagreement with
the exactly known behavior in the multi--channel case.  The mean field
theory can be shown to be exact for $M=1$ in the limit $N \rightarrow
\infty$ for a model in which the constraint is softened to be $Q =
N/2$. However, for finite $N$ i
t is known that the fluctuations in the phase of the complex
expectation value $\langle b_\mu \rangle$ are divergent and lead to the
suppression of $\langle b_\mu \rangle$ to zero.  This is true in the cartesian
gauge, whereas in the radial gauge the phase fluctuations may be shown
to cancel at least in lowest order. It has not been possible
to connect the mean field solution, an apparently reasonable
description at low temperatures and for $M = 1$, to the high temperature 
behavior $(T \gg T_K)$, dominated by logarithmic temperature dependence, 
in a continuous way \cite{readnewns.88}.  
Therefore, it seems that the slave boson mean field
solution does not offer a good starting point even for only a
qualitatively correct description of quantum impurity models.

\subsection{1/N expansion vs.~self-consistent formulation}
The critical judgement of mean field theory is corroborated by the
results of a straightforward $1/N$-expansion in the single channel
case, keeping the exact constraint, and not allowing for a finite
bose field expectation value \cite{kuroda.88}.  
Within this scheme the exact
behavior of the thermodynamic quantities (known from the Bethe ansatz
solution) at low temperatures as well as high temperatures is recovered
to the considered order in $1/N$.  Also, the exact auxiliary particle
exponents $\alpha _{f,b}$ are reproduced in order $1/N$, using a
plausible exponentiation scheme \cite{kuroda.97}.

In addition, dynamical quantities like the d-electron spectral
function and transport coefficients can be calculated exactly to a
desired order in $1/N$, within this approach.  However, as clear-cut
and economical this method may be, it does have serious limitations.
For once, the experimentally most relevant case of $N=2$ or somewhat
larger is not accessible in $1/N$ expansion. Secondly, non-Fermi
liquid behavior, being necessarily non-perturbative in $1/N$, cannot
be dealt with in a controlled way on the basis of a $1/N$-expansion.
To access these latter two regimes, a new approach nonperturbative in
$1/N$ is necessary.
\begin{figure}
\vspace*{-0cm}
\hspace*{0cm}{\psfig{figure=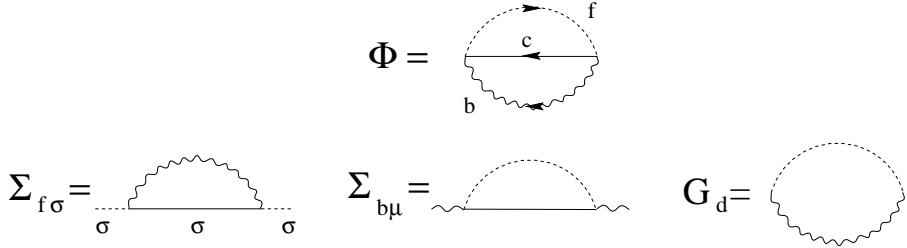,width=\textwidth}}
\caption{
Diagrammatic representation of the generating functional $\Phi $ 
of the NCA. Also shown are the pseudoparticle selfenergies and the
local electron Green's function derived from $\Phi$, Eqs.~(19)--(21).
Throughout this article, 
dashed, wavy and solid lines represent fermion, boson, and
conduction electron lines, respectively. In the diagram for 
$\Sigma _{f\sigma}$ the spin labels are shown explicitly to demonstrate
that there are no coherent spin fluctuations taken into account.
\label{NCA}}
\end{figure}

We conjecture that this new approach is gauge invariant many-body
theory of pseudofermions and slave bosons. As long as gauge symmetry
violating objects such as Bose field expectation values or fermion
pair correlation functions do not appear in the theory, gauge
invariance of physical quantities can be guaranteed in suitably chosen
approximations by the proper match of pseudofermion and slave boson
properties, without introducing an additional gauge field.  This
requires the use of conserving approximations \cite{kadanoff.61}, 
derived from a
Luttinger-Ward functional $\Phi$.  $\Phi$ consists of all vacuum
skeleton diagrams built out of fully renormalized Green's functions
$G_{b,f,c}$ and the bare vertex $V$.  The self-energies
$\Sigma_{b,f,c}$ are obtained by taking the functional derivative of
$\Phi$ with respect to the corresponding Green's function (cutting the
Green's function line in each diagram in all possible ways),
\begin{equation}
\Sigma_{b,f,c} = \delta \Phi/\delta G_{b,f,c}.
\label{fderiv}
\end{equation}
Irreducible vertex functions, figuring as integral kernels in
two-particle Bethe-Salpeter equations, are generated by second order
derivatives of $\Phi$.

The choice of diagrams for $\Phi$ defines a given approximation. It
should be dictated by the dominant physical processes and by
expansion in a small parameter, if available.  As noted before, in the
present context, we may take the hybridization $V$ to be a small
quantity (dimensionless parameter $N_oV$).  This suggests to start
with the lowest order (in $V$) diagram of $\Phi$, which is second
order (see Fig.~\ref{NCA}).  The self-energies generated from this obey after
projection the following equations of self-consistent second order
perturbation theory \cite{costi.96}
\begin{eqnarray}
\Sigma_{f\sigma}(\omega-i0)&=&V^2\sum _{\mu}\int {\rm d}\varepsilon\,
               [1-f(\varepsilon )]
              A_{c\mu\sigma}^0(\varepsilon)G_{b\mu}(\omega -\varepsilon -i0)
              \label{sigf}\\
\Sigma_{b\mu}(\omega -i0)&=&V^2\sum _{\sigma}\int {\rm d}\varepsilon\,
              f(\varepsilon )
              G_{f\sigma}(\omega +\varepsilon -i0)A_{c\mu\sigma}^0(\varepsilon)
              \label{sigb}\\
G_{d\mu\sigma}(\omega -i0)&=& 
\int {\rm d}\varepsilon \,  
             {\rm e}^{-\beta\varepsilon}
         [ G_{f\sigma}(\omega +\varepsilon -i0){\cal A }_{b\mu}(\varepsilon )
          \nonumber\\
    &\ &\hspace*{1.8cm}-{\cal A}_{f\sigma}(\varepsilon )
                   G_{b\mu}(\varepsilon -\omega +i0) ],
              \label{gd}
\end{eqnarray}
where $A_{c\mu\sigma}^0=1/\pi\ {\rm Im}G_{c\mu\sigma}^0$ is the
(free) conduction electron density of states and
$f(\varepsilon )=1/({\rm exp}(-\beta\varepsilon )+1)$ denotes the Fermi 
distribution function. Together with the expressions 
(\ref{green}), (\ref{green0}) for the Green's functions,
Eqs. (\ref{sigf})--(\ref{gd}) form a set of self-consistent equations for
$\Sigma_{b,f,c}$, comprised of all diagrams without any crossing
propagator lines and, thus, known as the
``non--crossing approximation'',  in short NCA.

At zero temperature and for low frequencies 
Eqs. (\ref{sigf}) and (\ref{sigb}) may
be converted into a set of linear differential equations for $G_f$ and
$G_b$ \cite{muha.84}, 
which allow to find the infrared exponents as $\alpha_f =
\frac{M}{M+N}$; $\alpha_b = \frac{N}{M+N}$, independent of $n_d$.  For
the single channel case these exponents do not agree with the exact
exponents derived in section 3. This indicates that the NCA is
not capable of recovering the local Fermi liquid behavior for $M=1$.
A numerical evaluation of the $d$-electron Green's function, which 
is given by the local self-energy $\Sigma_c$ divided by $V^2$ and 
hence is given by the boson-fermion bubble within NCA 
(Fig.~\ref{NCA}), shows indeed 
a spurious singularity at the Fermi energy \cite{costi.96}.  
The NCA performs somewhat better in the multi--channel case, 
where the exponents
$\alpha_f$ and $\alpha_b$ yield the correct non--Fermi liquid exponents
of physical quantities as known from the Bethe ansatz 
solution \cite{wiegmann.83} and conformal field 
theory \cite{affleck.91}. However, the specific heat and
the residual entropy are not given correctly in NCA.  Also, the
limiting low temperature scaling laws for the thermodynamic
quantities are attained only at temperatures substantially below
$T_K$, in disagreement with the exact Bethe ansatz solution.

\section{Conserving T-matrix approximation}
\begin{figure}
\vspace*{-0cm}
\centerline{\psfig{figure=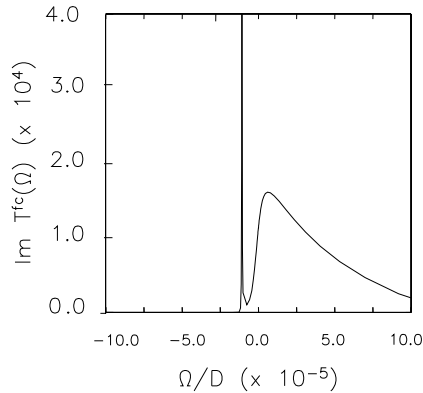,width=5.5cm}}
\caption{ 
Imaginary part of the conduction electron--pseudofermion 
T--matrix $T^{(cf)}$ as a function of the
COM frequency $\Omega$ for the single--channel case $M=1$, $N=2$, 
evaluated by inserting NCA solutions for the
intermediate state propagators ($E_d=-0.67 D$, $\Gamma = 0.15 D$,
$T=4\cdot 10^{-3}T_K$).
The contribution from the pole positioned at a negative frequency
$\Omega = \Omega _{cf} \simeq -T_K$ (compare text) is clearly seen.
\label{tmatpole0}}
\end{figure}

In order to eliminate the shortcomings of the NCA mentioned above, 
the guiding principle should be to find 
contributions to the vertex functions which renormalize the
auxiliary particle threshold exponents to their correct values,
since this is a necessary condition for the description of 
FL and non--FL behavior, as discussed in section 3. Furthermore,
it is instructive to realize that in NCA any coherent spin flip
and charge transfer processes are neglected, as can be seen 
explicitly from Eqs.~(\ref{sigf}), (\ref{sigb}) or from 
Fig.~\ref{NCA}. These processes are known to be responsible for the
quantum coherent collective behavior of the Anderson impurity
complex below $T_K$. The existence of collective excitations
in general is reflected in a singular behavior of the 
corresponding two--particle vertex functions. In view of the
tendency of Kondo systems to form a collective spin singlet
state, we are here interested in the spin singlet channel
of the pseudofermion--conduction electron vertex function and
in the slave boson--conduction electron vertex function.
It may be shown by power counting arguments that there are no 
corrections to the NCA exponents in any finite order of
perturbation theory \cite{coxruck.93}. Thus, we are led to 
search for singularities in the aforementioned vertex functions
arising from an infinite resummation of terms.   
\begin{figure}
\vspace*{-0cm}
\hspace*{0.65cm}{\psfig{figure=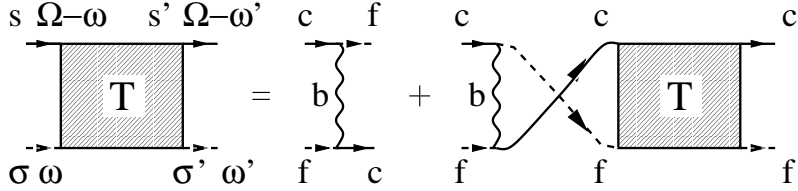,width=10.5cm}}
\caption{
Diagrammatic representation of the Bethe--Salpeter equation for the
conduction electron--pseudofermion
T-matrix $T^{(cf)}$, Eq.~(22). The conduction electron-slave boson 
T-matrix $T^{(cb)}$ is obtained by interchanging $f \leftrightarrow 
b^{\dag}$.
\label{tmat}}
\end{figure}
From the preceding discussion it is natural to perform a partial 
resummation of those terms which, at each order in the 
hybridization $V$, contain the maximum number of spin flip or charge 
fluctuation processes,
respectively. This amounts to calculating the 
conduction electron--pseudofermion vertex function in
the ``ladder'' approximation definied in Fig.~\ref{tmat}, where the 
irreducible vertex is given by $V^2G_b$. In analogy to similar 
resummations
for an interacting one--component Fermi system, we call the total
c--f vertex function T--matrix $T^{(cf)}$. 
The Bethe--Salpeter equation for $T^{(cf)}$ reads (Fig.~\ref{tmat}),
\begin{eqnarray}
T^{(cf)}_{s\sigma,s'\sigma '}(i\omega _n, i\omega _n ', i\Omega ) =
&-&V^2G_b(i\omega _n + i\omega _n ' - i\Omega ) \delta _{s\sigma'}\delta _{s'\sigma}
\nonumber\\
&+&V^2T\sum _{\omega _n''}G_b(i\omega _n + i\omega _n '' - i\Omega ) \times 
\label{tmateq}\\
&\ &G_{f\sigma}(i\omega _n'') \ G_{cs}(-i\omega _n ''+i\Omega )\ 
T^{(cf)}_{\sigma s,s'\sigma '}(i\omega _n '', i\omega _n ', i\Omega ). \nonumber
\end{eqnarray}\noindent
A similar  integral equation holds for the charge fluctuation T--matrix
$T^{(cb)}$. Inserting NCA Green's functions for the intermediate state
propagators of Eq. (\ref{tmateq}) and solving it numerically, 
we find at low
temperatures and in the Kondo regime $(n_d \buildrel >\over\sim  0.7)$ 
a pole of $T^{(cf)}$ in the singlet channel as a function of the 
center--of--mass
(COM) frequency $\Omega$, at a frequency which scales with the Kondo
temperature, $\Omega = \Omega_{cf} \simeq - T_K$. 
This is shown in Fig.~\ref{tmatpole0}. The threshold behavior of
the imaginary part of $T^{(cf)}$ as a function of $\Omega$ with
vanishing spectral weight at negative frequencies and temperature 
$T=0$ is clearly seen. In addition, a very sharp structure appears,
whose broadening is found to vanish as the temperature tends to zero, 
indicative of a pole in $T^{(cf)}$ at the {\it real} frequency
$\Omega _{cf}$, i.e.~the tendency to form a collective singlet 
state between the conduction electrons and the localized spin.   
Similarly, the corresponding $T$-matrix $T^{(cb)}$ in the conduction
electron--slave boson channel, evaluated within the analogous 
approximation,
develops a pole at negative values of $\Omega$ in the empty orbital
regime $(n_d \buildrel <\over\sim 0.3)$.  In the mixed valence 
regime ($n_d \simeq 0.5)$ the poles in both $T^{(cf)}$ 
and $T^{(cb)}$ coexist.
The appearance of poles in the two-particle vertex functions $T^{(cf)}$
and $T^{(cb)}$, which signals the formation of collective states, may
be expected to influence the behavior of the system in a major way.

\begin{figure}
\vspace*{-0cm}
\hspace*{1.5cm}{\psfig{figure=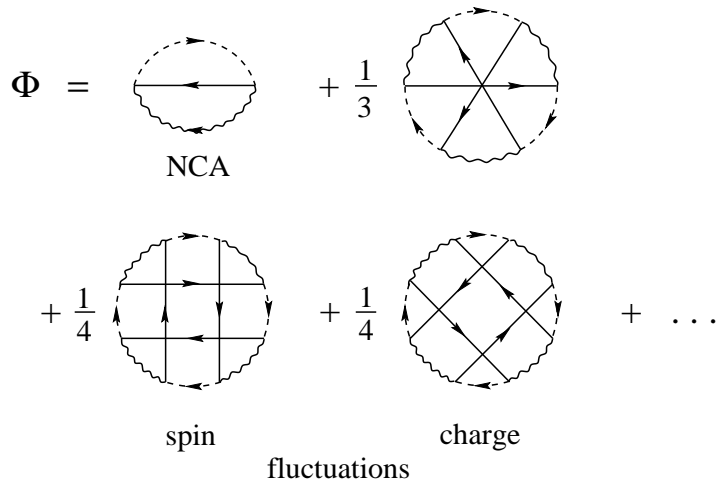,width=9.4cm}}
\caption{ 
Diagrammatic representation of the
Luttinger--Ward functional generating the conserving
T--matrix approximation (CTMA). The terms with the conduction electron
lines running
clockwise (labelled ``spin fluctuations'') generate the T-matrix 
$T^{(cf)}$, while the terms with the conduction electron lines
running counter--clockwise (labelled ``charge fluctuations'')
generate the T-matrix $T^{(cb)}$.
\label{CTMA}}
\end{figure}

On the level of approximation considered so far, the description is
not yet consistent: In the limit of zero temperature the spectral
weight of $T^{(cf)}$ and $T^{(cb)}$ at negative frequencies $\Omega$
should be strictly zero (threshold property). Nonvanishing 
spectral weight at $\Omega < 0$ like   
a pole contribution for negative $\Omega$ in
$T^{(cf)}$ or $T^{(cb)}$ would lead to a diverging contribution to the
self-energy, which is unphysical. However, 
recall that a minimum requirement on the approximation used is the
conservation of gauge symmetry. This requirement is not met when the
integral kernel of the $T$-matrix equation is approximated by the NCA
result. Rather, the approximation should be generated
from a Luttinger--Ward functional. The corresponding generating 
functional is shown in Fig.~\ref{CTMA}. It is defined as the 
infinite series of all vacuum skeleton diagrams which consist
of a single ring of auxiliary particle propagators, where each
conduction electron line spans at most two hybridization vertices.
The first diagram of this series corresponds to NCA
(Fig.~\ref{NCA}). The diagram containing two boson lines is 
excluded since it is not a skeleton. Although the spirit of the
present theory is different from a large $N$ expansion, it should be
noted that the sum of the $\Phi$ diagrams containing up to four 
boson lines includes all terms of a $1/N$ expansion up 
to $O(1/N^2)$ \cite{anders.94}. By functional differentiation
with respect to the conduction electron Green's function and
the pseudofermion or the slave boson propagator, respectively,
the shown $\Phi$ functional generates the ladder approximations
$T^{(cf)}$, $T^{(cb)}$ 
for the total conduction electron--pseudofermion vertex
function (Fig.~\ref{tmat}) and for the total conduction 
electron--slave boson vertex function. The auxiliary particle
self--energies are obtained in the conserving scheme as the 
functional derivatives of $\Phi$ with respect to $G_f$ or
$G_b$, respectively (Eq.~(\ref{fderiv})). This defines a set 
of self--consistency equations, which we term 
conserving T--matrix approximation (CTMA), 
where the self-energies are given as  
nonlinear and nonlocal (in time) functionals of the Green's functions,
while the Green's functions are in turn expressed in terms of the
self--energies. The solution of these equations 
requires that the T--matrices  have vanishing 
spectral weight at negative COM frequencies $\Omega$. Indeed, the
numerical evaluation shows
that the poles of $T^{(cf)}$ and $T^{(cb)}$ are shifted to 
$\Omega = 0$ by self--consistency, where they merge with the
continuous spectral weight present for $\Omega >0$, thus renormalizing
the threshold exponents of the auxiliary spectral functions.

The self-consistent solutions are obtained by first solving the
linear Bethe--Salpeter equations for the T--matrices
by matrix inversion, computing the auxiliary particle 
self--energies from $T^{(cf)}$ and $T^{(cb)}$, and then constructing
the fermion and boson Green's functions from the respective 
self--energies. This process is iterated until convergence is reached.  
We have obtained reliable solutions  
down to temperatures of the order of at least $10^{-2} T_K$
both for the single-channel and for the two-channel Anderson
model. Note that $T_K\rightarrow 0$ in the Kondo limit; 
in the mixed valence and empty impurity regimes, significantly lower
temperatures may be reached, compared to the low temperature scale of
the model. 

\begin{figure}
\hspace*{0cm}{\psfig{figure=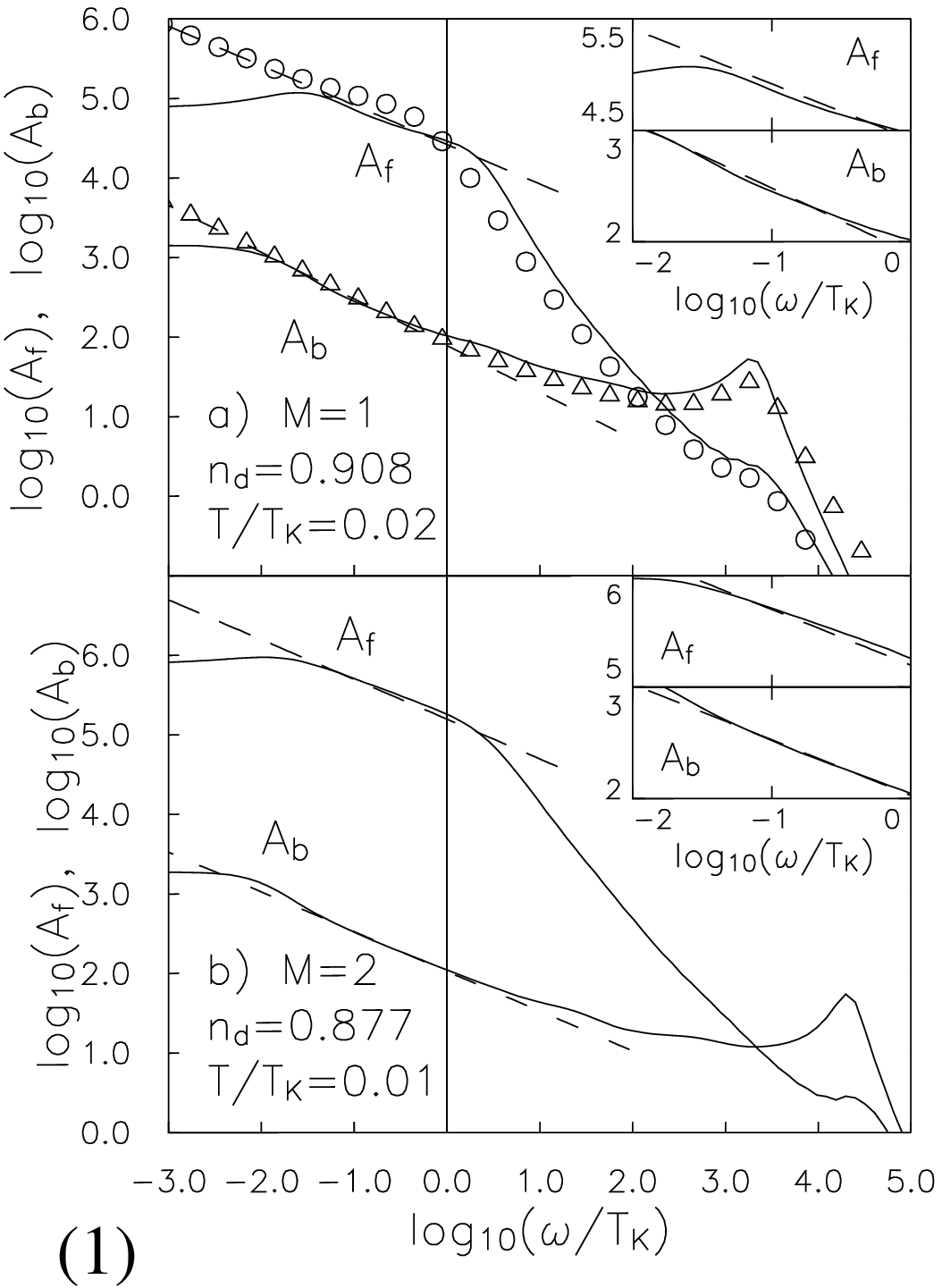,width=6cm}}
\hfill{\psfig{figure=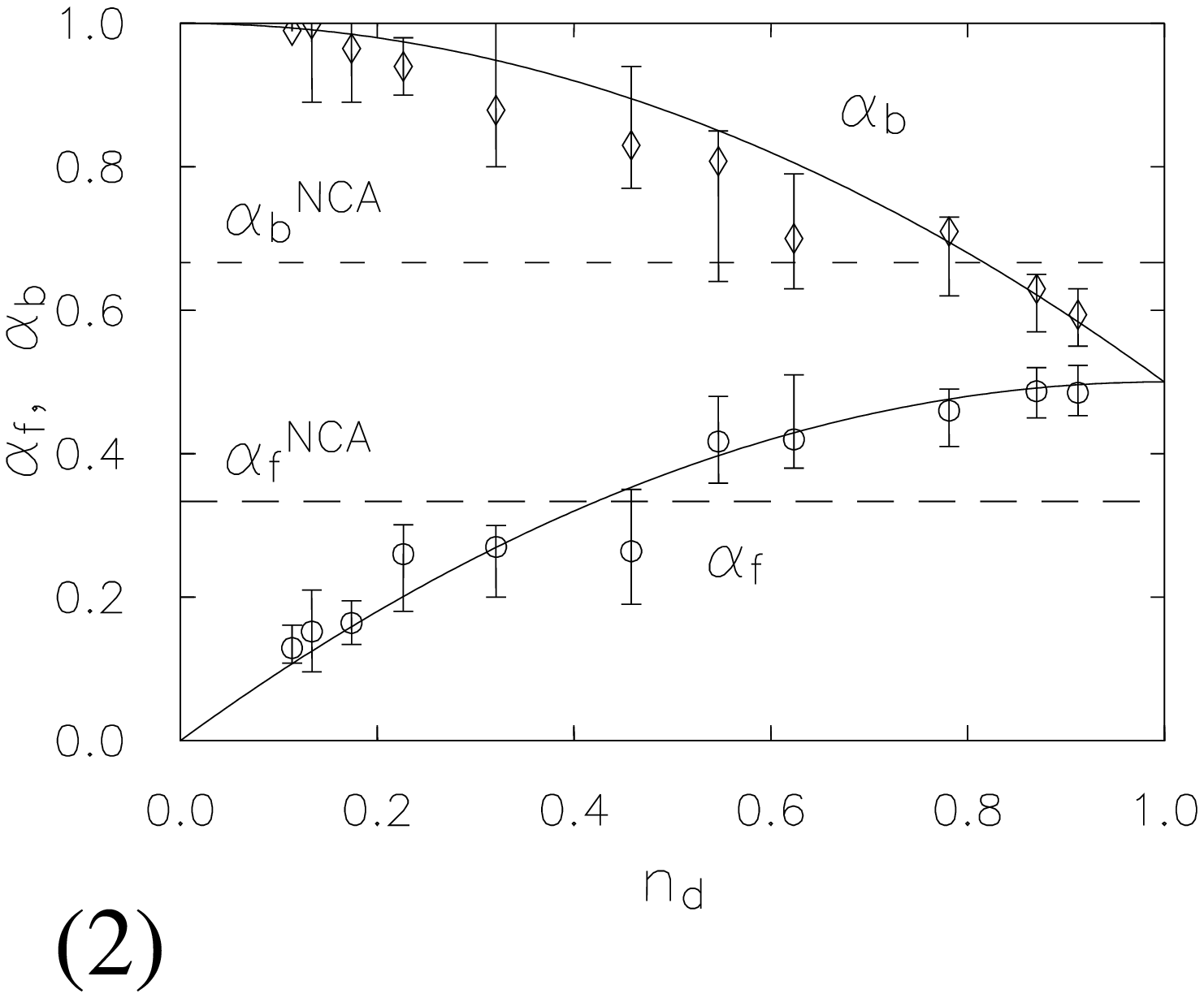,width=5.8cm}}
\caption{ 
(1) Pseudofermion and slave boson spectral functions $A_f$ and $A_b$
in the Kondo regime ($N=2$; 
$E_d=-0.05$, $\Gamma =0.01$ in units of the half--bandwidth $D$), 
for a) the single--channel ($M=1$) and b) the
multi--channel ($M=2$) case.
In a) the symbols represent the results of NRG for the same 
parameter set,
$T=0$. The slopes of the dashed lines indicate the exact
threshold exponents as derived in section 3 for $M=1$ and as given by 
conformal field theory 
for $M=2$. The insets show magnified power law regions. \ 
(2) CTMA results (symbols with error bars) for the threshold 
exponents $\alpha _f$ and $\alpha _b$ of $A_f$ and $A_b$, $N=2$, $M=1$. 
Solid lines: exact values (section 3), dashed lines: NCA results (section 4.2).
\label{spectralfb}}
\end{figure}

As shown in Fig.~\ref{spectralfb} (1) a), 
the auxiliary particle spectral functions obtained from CTMA 
\cite{kroha.97} are in 
good agreement with the results of a numerical renormalization 
group (NRG) calculation \cite{costi.94}
(zero temperature results), given the uncertainties in the 
NRG at higher frequencies. Typical behavior in the Kondo regime 
is recovered: a broadened peak in $A_b$ at $\omega\simeq |E_d|$, 
representing the hybridizing $d$--level and a structure 
in $A_f$ at $\omega \simeq T_K$. Both functions display power law 
behavior at frequencies
below $T_K$, which at finite $T$ is cut off at the scale $\omega
\simeq T$. The exponents extracted from the frequency range
$T<\omega<T_K$ of our finite $T$ results  
compare well with the exact result also shown
(see insets of Fig.~\ref{spectralfb} ($1 a$)). 
A similar analysis has been 
performed for a number of parameter sets spanning the complete range of
$d$--level occupation numbers $n_d$. The extracted power law exponents
are shown in Fig.~\ref{spectralfb} (2), 
together with error bars estimated 
from the finite frequency ranges over which the fit was made.
The comparatively large error bars
in the mixed valence regime arise because here spin flip and
charge fluctuation processes, described by the poles in $T^{(cf)}$ and
$T^{(cb)}$, respectively, are of equal importance, impeding the 
convergence of the numerical procedure. In this light, the agreement
with the exact results (solid curves) is very good, 
the exact value lying within the error bars or very close 
in each case. 

In the multi--channel channel case ($N\geq 2$, $M\geq N$) NCA has been
shown \cite{coxruck.93} 
to reproduce asymptotically the correct threshold 
exponents, $\alpha _f = M/(M+N)$, $\alpha _b = N/(M+N)$,
in the Kondo limit. Calculating the
T--matrices using NCA Green's functions
(as discussed in the single--channel case) we find again
a pole in the singlet channel of $T^{(cf)}$. However, the weight of
the pole vanishes in the Kondo limit $n_d \rightarrow 1$. As a
result, the CTMA does not renormalize the NCA exponents 
in the Kondo limit of the two--channel model, i.e.~the threshold 
exponents obtained from the CTMA solutions are very close to the
exact ones, $\alpha _f =1/2$, $\alpha _b=1/2$, as shown in 
Fig.~\ref{spectralfb} (1) b). 
  
\begin{figure}
\vspace*{-0cm}
\hspace*{1cm}{\psfig{figure=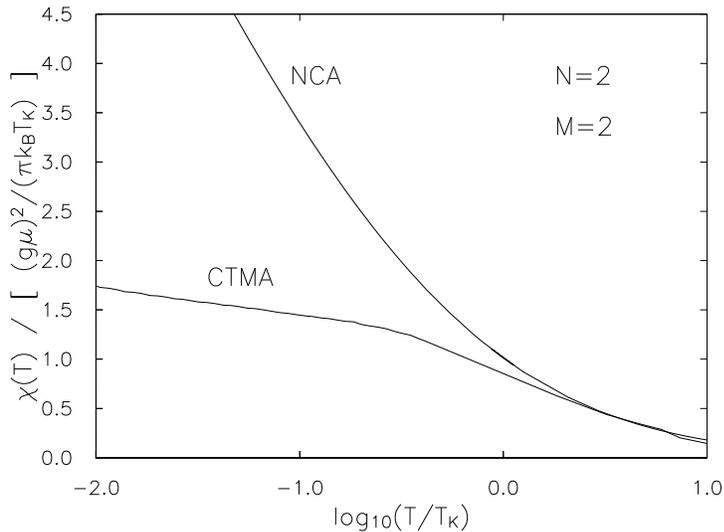,width=9.5cm}}
\caption{
Static susceptibility of the two--channel Anderson impurity model:
CTMA and NCA results ($E_d=-0.8D$, $\Gamma = 0.1D$, Land\'e factor $g=2$).
\label{suscM=2}}
\end{figure}

The agreement of the CTMA exponents with their exact values
in the Kondo, mixed valence and empty impurity regimes of the
single--channel model and in the Kondo regime of the
two--channel model  may be taken as
evidence that the T--matrix approximation correctly 
describes both the FL and the non--FL
regimes of the SU(N)$\times$SU(M) Anderson model (N=2, M=1,2).
Therefore, we expect the CTMA to correctly describe  
physically observable quantities of the SU(N)$\times$SU(M) Anderson 
impurity model as well.
In Fig.~\ref{suscM=2} we show
the static spin susceptibility $\chi$ of the two--channel Anderson 
model in the Kondo regime. It is seen that CTMA correctly reproduces
the exact \cite{wiegmann.83} 
logarithmic temperature dependence below the Kondo scale $T_K$.
In contrast, the NCA solution recovers the logarithmic behavior
only far below $T_K$. Other physical quantities will be
calculated for the Anderson model in forthcoming work.

\section{Conclusion}
We have presented a novel technique
to describe correlated quantum impurity systems with strong onsite
repulsion, which is based on 
a gauge invariant formulation of the auxiliary boson method.
This technique allows for the first time to describe physical
quantities, like the magnetic susceptibility, over the complete
temperature range, including the crossover to the 
correlated many--body state at the lowest temperatures. 
The numerical effort involved in the evaluations of
the multiple integrals is manageable, considering that one complete 
iteration within the self--consistent scheme has been done on a parallel
computer within approximately 5--10 s CPU time.   
As a standard diagram technique this method has the potential to be 
applicable to problems of correlated systems on a lattice as well,
while keeping the full dynamics of the pseudofermion and slave boson fields. 

%\section*{Acknowledgments}
We wish to thank S.~B\"ocker, 
T.A.~Costi, A.~Rosch, and A.~Ruckenstein for stimulating discussions.
S.~B\"ocker has performed part of the numerical solutions. 
This work is supported by DFG through SFB 195 and by the
Hochlei\-stungsrechenzentrum J\"ulich through a grant of computer 
time on a Cray T3E parallel computer.

\end{document}